\documentstyle[prl,aps,multicol,psfig]{revtex} \tighten
\begin{document}
\draft
\title{New quantum phases in a one-dimensional Josephson array}
\author{A.I. Larkin$^{1,2}$, L.I. Glazman$^{1}$}  
\address{
$^1$Theoretical Physics Institute, University of Minnesota, 
 Minneapolis, MN 55455 \\
$^2$L.D. Landau Institute for Theoretical Physics, 117940 Moscow, Russia}
\date{Draft of \today}
\maketitle
\begin{abstract}
We examine the phase diagram of an ordered one-dimensional Josephson
array of small grains. The average grain charge in such a system can be
tuned by means of gate voltage. At small grain-to-grain conductance,
this system is strongly correlated because of the charge discreteness
constraint (Coulomb blockade). At the gate voltages in the vicinity of the charge
degeneracy points, we find new phases equivalent to a commensurate charge density wave
and to a repulsive Luttinger liquid. The existence of these phases can be
probed through a special dependence of the Josephson current on the gate voltage.
\end{abstract}

\begin{multicols}{2} 

In the physics of phase transitions, the special class of transitions
occuring at zero temperature have attracted significant attention in the
recent years\cite{Girvin}. At zero temperature, long-range order may be
destroyed by quantum fluctuations. In some systems, the magnitude of
these fluctuations is controlled by an external parameter. Therefore such
a system may undergo transitions between the phases with qualitatively
different behavior of long-range correlations. In contrast to the
conventional finite-temperature transitions, the quantum transition at
$T=0$ can occur even in a one-dimensional system. For example, as it was
predicted by Bradley and Doniach\cite{Doniach}, a one-dimensional array
of superconducting grains linked by Josephson junctions may behave as a
superconductor or insulator depending on the ratio of charging and
Josephson energies. Recently a special sample design allowing one to
vary continuously the strength of inter-grain Josephson junctions was
developed\cite{Haviland1} and an attempt to examine the vicinity of the
quantum phase transition was undertaken\cite{Haviland2}.

Paper\cite{Doniach} and the majority of later theoretical papers devoted
to the one-dimensional Josephson junction arrays,
assumed that the grains of an array are tuned to the point of
particle-hole symmetry. In this case, the transition from a
superconductor to Mott insulator is of the Berezinskii-Kosterlitz-Thouless (BKT)
type\cite{Doniach}. Within a broader framework of quantum phase
transitions in a system of interacting bosons\cite{Fisher4}, it became clear that
breaking the particle-hole symmetry immediately changes the universality class of the
transition.  

Much of the physics of interacting bosons can be applied directly to the
Josephson junction arrays. For an array, a boson is replaced by a Cooper
pair. Breaking the particle-hole symmetry means simply applying some
voltage $V_g$ to a gate which is electrostatically coupled to all grains
of the array. Breaking the symmetry leads to a symmetric in $V_g$ shift of the
superconductor-insulator transition. Recently\cite{Allen} some experimental evidence
was found for such a symmetric shift for two-dimensional systems.

In this paper, we found new, compared to the predicted in\cite{Fisher4}, phases of a
one-dimensional Josephson array. These appear due to the effective repulsion between
Cooper pairs at different sites of the array. In the plane of controlling parameters,
which are the Josephson energy $E_J$ and voltage
$V_g$, these phases occupy the vicinities of the charge degeneracy points. At $E_J=0$
the effective repulsion leads to a period-doubling transition. It separates the known
Mott phase with homogeneously distributed Cooper pairs, from a new phase with
alternate occupation of the grains (``Neel phase''). As $E_J$ increases, the regions
of Mott and Neel phases separate from each other, giving way to another, intermediate
phase. This phase is equivalent to a Luttinger liquid with repulsion. Further
increase of the Josephson coupling brings the system to  a transition into the
superconducting state (Luttinger liquid with attraction). We found the corresponding
transition line on the phase diagram $\{E_J, V_g\}$, see Fig.~1. This line can be
identified with an abrupt change in the behavior of the Josephson current through a
chain containing one weaker link. 

We also found the critical exponents for the
transition line between the Luttinger liquid and Mott phase.

To start with, we consider a one-dimensional Josephson array in the
absence of Cooper pair repulsion. The simplest Hamiltonian that allows for the
superconductor-insulator phase transition, has the form:
\begin{eqnarray}
H=H_C+H_J,\quad
H_C=\frac{E_C}{2}\sum_{i}\left(\frac{\partial}{i\partial\phi_i}-
\frac{{\cal N}}{2}\right)^2, \nonumber\\
H_J=-E_J\sum_i\cos(\phi_{i+1}-\phi_i).\label{H}
\end{eqnarray}
In comparison with\cite{Doniach,Anderson,Efetov}, Hamiltonian (\ref{H})
accounts for the influence of the gate voltage, $e{\cal N}\propto V_g$ (in the
absence of charge quantization, $e{\cal N}$ is the average grain charge
induced by the gate); $E_C$ is the charging energy of a grain. 

The two parts of the Hamiltonian (\ref{H}) do not commute with each
other, leading to quantum fluctuations of phase $\phi_i$ of the
superconducting order parameter in each grain. To start with, we
consider the domain of weak Josephson coupling, $E_J\ll E_C$, where
the fluctuations are strong. In the limit $E_J\to 0$, the Josephson
array is a Mott insulator at almost any ${\cal N}$. The exceptions are
the discrete points ${\cal N}=2n+1$, where a grain with charges $2ne$ and
$2(n+1)e$ has the same energies. Josephson tunneling lifts this
degeneracy. To analyze the properties of the array in the vicinity of
the degeneracy points, we can project the Hamiltonian (\ref{H}) on the
subspace of states with charges of each grain confined to the values
$2ne$ and $2(n+1)e$ only. In the lowest order in $E_J$ the projected
Hamiltonian has the form:
\begin{eqnarray}
H_C^{(0)}=
\frac{E_C}{2}\sum_{i}\left(\sigma^z_i-h\right)^2,
\nonumber\\
H_J^{(0)}=\frac{E_J}{2}\sum_i\left(\sigma^+_i\sigma^-_{i+1}+
\sigma^-_i\sigma^+_{i+1}\right),
\label{H1}
\end{eqnarray}
where $\sigma^z,\,\sigma^+,\,\sigma^-$ are the Pauli matrices, and the
``magnetic field'' $h=({\cal N}-2n-1)/2$ allows for tuning the system to a degeneracy
point by means of the gate voltage. The Hamiltonian (\ref{H1})  can be converted to
the conventional tight-binding Hamiltonian for free fermions by means of the
Jordan-Wigner transformation, and then diagonalized\cite{Mattis} in the plane wave
representation:
\begin{equation}
H_0=\sum_k\varepsilon_k a_k^\dagger a_k, \quad 
\varepsilon_k=E_J\cos k - E_Ch.
\label{H2}
\end{equation}
The fermion band is completely full or totally empty, if 
$|h|> E_J/E_C$. This condition defines the location of
the Mott phase in the plane of dimensionless parameters $\{{\cal N},\,
E_J/E_C\}$. The Mott gap, 
$\epsilon_M=\left|E_C|h|-E_J\right|$, coincides
with the lowest possible excitation energy, and vanishes at the phase
boundary. Clearly, the phase diagram is periodic in ${\cal N}$ with period 2.
This allows us to consider further only a ``strip'' of the phase
diagram, $0\leq{\cal N}\leq 2$ which corresponds to $n=0$ in the above formulas.

The lowest order in $E_J$ we used is sufficient to find the
characteristics of the Mott phase. The complementary phase in
this approximation is equivalent to a non-interacting Fermi gas. It is
known however, that even a weak interaction is relevant in one dimension,
transforming the gas into a Luttinger liquid. There are two competing
mechanisms of interaction between the fermions. 

The inter-grain electrostatic
interaction leads to an effective repulsion between the Cooper pairs.
We will concentrate on the case of a relatively weak inter-grain
interaction, $E_z\ll E_C$, which allows us to use Hamiltonian
(\ref{H1}) as a starting point. In a realistic system with a gate, the
interaction between distant grains is screened out, and only the
nearest-neighbor term should be retained:
\begin{equation}
H_z=\sum_iE_z\sigma_i^z\sigma_{i+1}^z.
\label{Hz}
\end{equation}
On the other hand, the higher order expansion in $E_J/E_C$ leads to an effective
attraction between the Jordan-Wigner fermions.
To find the corresponding correction to the low-energy Hamiltonian (\ref{H1}), we
should take into account the virtual states with energies exceeding $E_C$. To do
this, we use the perturbation theory. In the second-order perturbation theory, the
effective Hamiltonian reduced to the subspace of charges $0$, $2$, takes the form:
\begin{equation}
H_2=PH_J(1-P)\frac{1}{\varepsilon - H_C}H_JP.
\label{HP}
\end{equation}
Here $P=\prod_i P_i$, and $P_i$ is the projection operator on the
subspace of states with charges $0$ and $2$ for each grain; energy
$\varepsilon$ belongs to the reduced energy band, and may be neglected in
comparison with $E_C$. To implement the projection procedure in
(\ref{HP}), we use the identities,
$P_i\exp(\pm\phi_i)(1-P_i)exp(\mp\phi_i)P_i=(1\mp\sigma_i^z)/2$, that
can be easily checked. Thus, the result of projection can be written in
terms of Pauli matrices,
\begin{equation}\!\!
H_2=-\frac{E_J^2}{4E_C}\sum_i
\left(\frac{3}{4}\sigma_i^z\sigma_{i+1}^z+
\sigma_{i+1}^+\sigma_{i-1}^-+\sigma_{i-1}^+\sigma_{i+1}^-\right)\!.
\label{HPsigma}
\end{equation}
The full Hamiltonian $H^{(0)}+H_z+H_2$ enables us to investigate the
phase diagram in the vicinity of the charge degeneracy line ($h\ll 1$).

We can neglect the interaction $H_2$, if $E_J^2/E_CE_z\ll 1$; in this
limit, our system with Hamiltonian $H^{(0)}+H_z$ is equivalent to a
one-dimensional anisotropic Heisenberg model in a magnetic field. It is
known\cite{Fowler} that  at $E_J<4E_z$ and relatively small field $|h|<h_1$, the
antiferromagnetic Izing interaction dominates the physics of the anisotropic
Hesenberg model, whereas in a large field, $|h|>h_2$, the chain is in
the ferromagnetic state.  Returning to the language of the initial chain of grains,
the ferromagnetic phase corresponds to Mott insulator. The low-field
Neel phase is a commensurate charge density wave state with period 2 (on
average, every second grain of the chain is occupied by a Cooper pair). In both
states, there is a gap for excitations. The region in between, $h_1<|h|<h_2$, is
occupied by Luttinger liquid. The width of the Luttinger liquid region depends on the
value of $E_J$. In the limiting case $E_J=0$ which corresponds to the Izing model,
this width shrinks to zero. At small $E_J$, the dependence of both
$h_1$ and $h_2$ is linear in $E_J$ with coefficients of opposite sign\cite{Fowler},
and the width of the Luttinger liquid region is proportional to $E_J$. The field $h_1$
turns zero for an isotropic antiferromagnet, $E_J=4E_C$. At larger $E_J$, the
Neel state is absent, and the liquid phase exists even at $h=0$, see Fig.~1.

With the further increase of $E_J$, the interaction $H_2$ becomes important. As we
will see, it alters the sign of interaction in the Luttinger liquid. To study the
interaction, we transform the spin Hamiltonian $H_z+H_2$ to the Jordan-Wigner
fermion representation. In addition to quadratic in $a$, $a^\dagger$ terms, this
Hamiltonian gives rise also to two-body interaction $V_{\rm int}$. The significant
part of (weak) interaction $V_{\rm int}$ that defines the properties of Luttinger
liquid, corresponds to the interaction between the left- and right-movers. Near the
one-dimensional Fermi ``surface''
$k=\pm k_F$ this part can be written as:
\begin{eqnarray}
&V&_{\rm int}=E_J\lambda\sum_{k_1,k_2,q}\!\sin^2k_F\,
a_{k_1}^\dagger a_{k_2}^\dagger a_{k_2+q} a_{k_1-q},
\label{Vint}\\
&\lambda&=\left(\frac{8E_z}{E_J}-\frac{7E_J}{2E_C}\right).\nonumber
\end{eqnarray}
Here $k_1\approx k_F$, $k_2\approx -k_F$, and $q\ll k_F$; the Fermi wavevector
varies with the gate voltage.

Interaction (\ref{Vint}) vanishes at the phase boundary of the Luttinger liquid with
Mott phase ($k_F\to 0$ or $k_F\to\pi$), regardless the relations between parameters
$E_z$, $E_J$, and $E_C$. Therefore, we can use the free-fermion model to study the
critical behavior of our system near the phase boundary. The divergent length scale,
$\xi (h)\sim 1/k_F$, characterizes the distance between free charges in the
conducting phase. Using Eq.~(\ref{H2}), we find $\xi\propto \delta^{-1/2}$, so that
the critical exponent $\nu$ equals $1/2$ (here $\delta=h_2-|h|$ is the distance to the
phase boundary). The characterisitic frequency scale $\Omega$ is given simply by the
Fermi energy. The corresponding exponent $z$, defined by the relation  $\Omega\sim
\delta^{z\nu}$, is $z=2$. Finally the superfluid density\cite{Fisher4},
$\rho_s\propto \delta^\zeta$, should be identified with the density of fermions
$\rho_s\sim \delta^{1/2}$, which determines $\zeta=1/2$.

Vanishing of the interaction at the phase boundary with Mott phase is a property
far more general than the result of perturbation theory (\ref{Vint}), and it
reflects the fact that we deal with {\em spinless} fermions. In the higher orders in
$E_J/E_C$ the dependence of $h_2$ on $E_J$ becomes nonlinear, and the spectrum of
fermions changes, but the critical exponents we found are not affected as long as
${\cal N}\neq 0,2$. The lines ${\cal N}=0,2$ are the lines of particle-hole symmetry.
Because of this symmetry, the system belongs to a different universality
class\cite{Fisher4}, and the transition to the Mott phase (which occurs at
$E_J/E_C\approx 4/\pi^2$) is of the BKT type.

Now we discuss the properties of the Josephson junctions chain in the Luttinger
liquid state, which low-energy properties are described by a quadratic
Lagrangian\cite{Efetov1,Haldane},
\begin{equation}
{\cal L}=\frac{g}{2\pi}[v(\partial_x\phi)^2+v^{-1}(\partial_\tau\phi)^2].
\label{L}
\end{equation}
Here $v$ is the velocity of the acoustic excitations in the system (plasmons in our
case), and $g$ determines the long-range behavior of all correlation
functions. The phenomenological constant $g$ can be calculated in terms of the
microscopic parameters in some limiting cases. 

{\narrowtext
\begin{figure}[h]
\vspace*{-1cm}
\hspace*{-1cm}
\psfig{figure=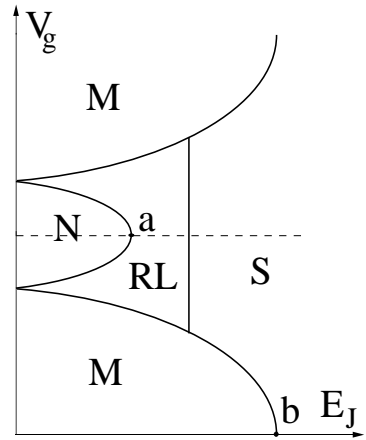,height=8cm}
\vspace{-1.5cm}
\caption{
Zero-temperature phases of the one-dimensional array of Josephson
junctions controlled by the gate voltage ($V_g$) and Josephson energy ($E_J$). In
addition to the known\protect\cite{Fisher4} Mott phase $M$ and superconducting
phase $S$ (which is equivalent to the attractive Luttinger liquid), we find the
commensurate charge density wave state $N$, and repulsive Luttinger liquid phase
$RL$. The Luttinger liquid parameter $g$ equals $1$ at the $RL$--$S$ boundary, and
also at the boundary of the Mott phase; $g=1/4$ at $RL$--$N$ boundary. The special
points on the phase transition line where the charge degeneracy occurs ($a$), and
where the particle-hole symmetry is observed ($b$), are multicritical points with
$g=1/2$ and $g=2$ respectively.} 
\vspace{-0.22cm}
\label{phases}
\end{figure}
}

Deep in the superconducting region, $E_J\gg E_C$, one can compare Eq.~(\ref{L}) with
the quadratic Lagrangian derived by expansion of (\ref{H}) in $\phi$, to
find $g\approx\pi(E_J/E_C)^{1/2}$. In the opposite limit of a relatively weak
Josephson coupling, $E_z\ll E_J\ll E_C$, constant $g$ can be found
with the help of (\ref{Vint}),  $g\approx 1-(\lambda/2\pi)\sin k_F$. In addition to
these limiting expressions for $g$, we know its value at the phase boundaries. We
already mentioned that at the boundary with the Mott phase the interaction is
vanishing, which means $g=1$. The exceptions are the special multicritical
points\cite{Fisher4}, where the boundary intersects with a particle-hole symmetry
line (point $b$ in Fig.~1). If point $b$ is crossed along the symmetry line, the
transition is of the BKT type with $g=2$. In the vicinity of these points,
$g$ rapidly varies from $g=1$ to $g=2$. With the help of Haldane's theory\cite
{Haldane1} for anizotropic $s=1/2$ Hezenberg model in a magnetic field, we find
$g=1/4$ for the boundary between the Luttinger liquid and Neel phases. The exceptions
are the multicritical points ($a$ in Fig.~1) of charge degeneracy, $h=0$, where
$g=1/2$. Similarly to the lines of particle-hole symmetry, the phase transition at
$h=0$ is of the BKT type, see, {\sl e.g.},\cite{Shankar}.

Different phases on the diagram Fig.~1 can be distinguished by the dependence of the
critical Josephson current on the length of the system $L$. Clearly, the gap in
the excitation spectrum existing in the insulating phases leads to an exponential
suppression of the critical current with $L$; the corresponding correlation length
diverges (with exponent $\nu$) at the transition line. 

To characterize the Luttinger liquid state, we consider a chain of junctions
connecting two massive superconducting leads, and containing just one especially weak
link, $E_w\ll E_J$. Deep in the superconducting state ($g\gg 1$) we can
neglect with the quantum fluctuations. The energy of the system consisting of $L$
junctions of the nominal strength $E_J$ and one weak link equals:
\begin{equation}
E(\phi)=\frac{E_J}{2}\frac{(\phi -\phi_w)^2}{L}-E_w\cos\phi_w.
\label{classical}
\end{equation}
Here $\phi$ is the phase difference applied to the leads, and $\phi_w$ is the phase
difference accross the weak link; the energy (\ref{classical}) should be minimized
with respect to $\phi_w$. Quantum fluctuations of phases $\phi_i$ at finite
$g$ lead to zero-point motion of $\phi_w$ around some average value $\bar{\phi}_w$,
which depends on the external phase $\phi$. Averaging over the fluctuations
$\delta\phi_w=\phi_w-\bar{\phi}_w$ results in the energy functional of the form
(\ref{classical}) with $E_w$ renormalized by a proper Debye-Waller factor, and
$\phi_w$ replaced by $\bar{\phi}_w$, see, {\sl e.g.},\cite{Hekking}. For a short array, the renormalized Josephson
coupling constant is $E_w^{\rm eff}\simeq E_w(1/L)^{1/g}$. Energy (\ref{classical})
corresponds to two inductors in series. Therefore, the response of the system to the
external phase is dominated by the weaker of two elements. If $g>1$ (weaker quantum
fluctuations), the renormalized energy $E_w^{\rm eff}$ decreases with $L$ slower than
the first term in Eq.~(\ref{classical}). Consequently, at $L$ exceeding the
crossover length, $L^*\simeq (E_J/E_w)^{g/(g-1)}$,
the energy (\ref{classical}) is dominated by the first term with $\phi_w=2\pi n$. The
Josephson current $\propto \partial E/\partial\phi$ has a sawtooth dependence on
phase, with the amplitude proportional to $E_J/L$. Thus, the effect of the weak link
is ``healed'' over the distance $L^*$. 	

If $g<1$, the weak link energy $E_w^{\rm eff}$ falls off faster than $1/L$, and this
link remains the weakest of the two inductors connected in series at any $L$;
consequently, $\bar{\phi}_w=\phi$. Therefore, ``healing'' does not occur, and at any
$L$ the chain behaves as a single Josephson junction of strength
$E_w^{\rm eff}$. The corresponding current-phase relation is $I\sim E_w
L^{1/g}\sin\phi$.

The line $g=1$ on the phase diagram may be viewed as a phase
transition\cite{transition} between the superconducting and insulating phases in the
following sense. In the limit $L\to\infty$, the inductance ${\cal L}$ of an array with
a single weak link is insensitive to the weak link if $g>1$ (superconducting state),
and ${\cal L}\propto L/E_J$. In the insulating state, the scaling of inductance with
$L$ is $g$-dependent, ${\cal L}\propto L^{1/g}/E_w$, with exponent
$1/g>1$\cite{Larkin}. 

To facilitate the quantitative analysis of the phase diagram, we considered the
limit $\Delta\gg E_C\gg E_z$. Eazing the first, and lifting the second of these
two restrictions does not alter the diagram qialitatively. Moreover, in a real
system with large inter-grain capacitance\cite{Haviland2} the energies
$E_C$ and $E_z$ are of the same order, and the domains of charge density wave and of
the repulsive Luttinger liquid are not small. The phase diagram changes drastically
only if the superconducting gap $\Delta$ becomes smaller than some critical value, at
which the grains may carry single electrons in the ground state. The charge density
wave state is destoyed then, and a phase appears with an odd charge on each grain.
For this phase, the Josephson current scales exponentially with $L$, but there is no
gap in the excitation spectrum due to the dense quasiparticle spectrum in each
superconducting grain. 

We assumed a perfect order in the system, which means there are no spatial variations
of the parameters of the Hamiltonian (\ref{H}). The validity of the above analysis
requires that the Josephson junctions chain is shorter than the mean free path for a
Cooper pair propagating along the chain.

In conclusion, we examined the zero-temperature phase diagram of a one-dimensional
chain of Josephson junctions. We have shown that accounting for the repulsion
between Cooper pairs occupying different grains leads to new phases, which are
equivalent to the phases of one-dimensional spinless fermions with repulsion on a
lattice. We have studied in detail the boundaries of the quantum phases.

We are grateful to I.L. Aleiner, F. Hekking, T. Giamarchi, S.M. Girvin, and 
D. Haviland for discussions and numerous references to papers
on related subjects. This work is supported by NSF Grant No. DMR-9423244.

\end{multicols}
\end{document}